\documentclass[preprint,showpacs,preprintnumbers,amsmath,amssymb]{revtex4}

\usepackage{graphicx}
\usepackage{graphics}
\begin{document}

\title{Crossover between 
L$\acute{\rm e}$vy and Gaussian 
regimes in first passage processes}
\author{Jun-ichi Inoue$^{1}$}
\email[e-mail: ]{j_inoue@complex.eng.hokudai.ac.jp}
\author{Naoya Sazuka$^2$}
\email[e-mail: ]{Naoya.Sazuka@jp.sony.com}
\affiliation{
$^1$Complex Systems Engineering, Graduate School of Information
Science and Technology, Hokkaido University, 
N14-W9, Kita-ku, Sapporo 060-0814, Japan \\
$^2$Sony Corporation, 
1-7-1 Konan Minato-ku, 
Tokyo 108-0075, Japan
}
\begin{abstract}
We propose a new approach to the problem of 
the first passage time. 
Our method is applicable  
not only to 
the Wiener process but also to 
the non-Gaussian L$\acute{\rm e}$vy flights 
or to more complicated stochastic 
processes whose distributions 
are stable. 
To show the usefulness of 
the method, we particularly focus on 
the first passage time problems 
in the truncated L$\acute{\rm e}$vy flights 
(the so-called KoBoL processes), 
in which the 
arbitrarily large tail 
of the L$\acute{\rm e}$vy distribution 
is cut off. 
We find that 
the asymptotic scaling law of the 
first passage time $t$ distribution 
changes from 
$t^{-(\alpha +1)/\alpha}$-law (non-Gaussian 
L$\acute{\rm e}$vy regime) to 
$t^{-3/2}$-law (Gaussian regime) 
at the crossover point. 
This result 
means that 
an ultra-slow convergence 
from the non-Gaussian L$\acute{\rm e}$vy 
regime to the Gaussian regime 
is observed not only in 
the distribution 
of the real time step for the 
truncated L$\acute{\rm e}$vy 
flight 
but also in the 
first passage time 
distribution of the flight. 
The nature of the 
crossover in the scaling laws 
and the scaling relation 
on the crossover point 
with respect to the effective 
cut-off length of the 
L$\acute{\rm e}$vy 
distribution are discussed. 
\end{abstract}

\maketitle
\section{Introduction}
The first passage process or 
the first passage time (FPT) problem 
deals with the 
event where 
a diffusing particle or a random-walker firstly 
reaches a specific site at a specific time \cite{Redner}. 
These FPT problems have been 
studied in various research fields, such as 
statistical physics, 
chemistry \cite{Kappen} and 
biological neuroscience \cite{Tuckwell,Tuckwell2}. 
In finance, several authors \cite{Simonsen,Raberto,Scalas,Kurihara,Sazuka,Sazuka2} have analysed tick-by-tick data of 
the US dollar/Japanese yen (USD/JPY) exchange rate and studied 
the FPT distribution for which the FPT is defined 
by the time that the rate firstly 
moves out from a given range.  

Among these studies, the USD/JPY exchange rates of the 
Sony Bank \cite{SonyBank} 
are reproduced from the market rates by using some 
{\it rate windows} with a 
width of $0.1$ yen \cite{Sazuka,Sazuka2}. 
That is, if the USD/JPY market rate changes by more than 
$0.1$ yen, 
the Sony Bank rate for USD/JPY is updated to 
the market rate otherwise it remains constant. 
In this sense, it is possible for us 
to say that the procedure for determining 
the USD/JPY exchange rate of the Sony Bank 
is essentially the first passage process. 
Despite many demands from 
various research fields 
and business in financial markets, 
one could obtain 
explicit analytical expressions or solutions of 
the FPT distribution only in very few cases. 
In addition, except for a few cases \cite{Rangarajan}, 
most of the analytical expressions 
are of the 
ordinary Wiener process (ordinary Brownian motion). 

Based on this fact, 
here we propose a new approach to the problem of 
the FPT or 
first passage processes. 
Our method is applicable  
not only to 
the Wiener process but also to 
the anomalous diffusion of 
the non-Gaussian L$\acute{\rm e}$vy flights 
or more complicated stochastic processes. 
To show the usefulness of 
our approach, we particularly focus on 
the FPT problems 
in the truncated L$\acute{\rm e}$vy flights 
\cite{Mantegna94,Mantegna94_2,Koponen}, 
in which the arbitrarily large tail 
of the L$\acute{\rm e}$vy distribution 
is cut off. 
Using the method, we find that 
the asymptotic scaling law of the 
FPT $t$ distribution 
changes from 
a $t^{-(\alpha +1)/\alpha}$-law (non-Gaussian 
L$\acute{\rm e}$vy regime) to 
a $t^{-3/2}$-law (Gaussian regime) 
at some crossover point. 
This fact  
means that the crossover between 
non-Gaussian L$\acute{\rm e}$vy 
and Gaussian regimes 
is observed not only in 
the distribution 
of the real-time 
step of the truncated L$\acute{\rm e}$vy flight, 
which was reported by Mantegna and 
Stanley \cite{Mantegna94}, 
but also in the FPT distribution of the flight. 
Moreover, 
we give a scaling relation on the 
crossover point with respect to 
the effective cut-off length 
of the 
L$\acute{\rm e}$vy distribution. 
The scaling 
relation enables us to predict the crossover point 
of the FPT distribution 
for a given truncated L$\acute{\rm e}$vy flight. 

This paper is organised as follows. 
In the next section, 
we explain general formalism of 
our method and 
apply it to 
the FPT problem for 
the Wiener process, 
for which the solution 
of the FPT distribution is well-known, 
in order to check the validity of our method. 
In Sec. III, 
we show that our method is widely useful for a class of 
stable stochastic 
processes. We derive the 
FPT distribution for 
L$\acute{\rm e}$vy flight which includes 
Gaussian and Lorentzian stochastic processes as its 
special cases. For each 
stable stochastic process, 
we discuss the scaling law of 
the FPT distribution in 
the asymptotic regime. 
The analytical results are confirmed by computer simulations. 
In Sec. IV, we apply our method to 
the FPT problem of the truncated L$\acute{\rm e}$vy flight and 
discuss the crossover in the scaling laws of 
the FPT distribution 
between non-Gaussian L$\acute{\rm e}$vy and 
Gaussian regimes. 
The last section is a summary. 
\section{General formalism}
The problem we deal with 
in this paper is defined as follows. 
Let us consider the 
stochastic process : 
$X_{k}: k=0,1,\cdots,T$. 
For this time series, the FPT 
$t$ is defined by 
$t =\min\{k \geq k_{0} \,;\, X_{k}=\pm \epsilon \}$. 
Then, our problem is to 
obtain the distribution of 
$t$, namely, the {\it first passage time 
distribution} $P(t)$. 
In other words, 
we evaluate the distribution of 
$t$, that is $P(t)$, which is defined 
as the 
survival probability 
that the time series $X_{k}$, starting from 
$k=k_{0}$ keeps 
staying within the range $[X_{k_{0}}-\epsilon, 
X_{k_{0}}+\epsilon]$ up to 
the time step $k_{0}+t$.
The problem we are dealing with is motivated 
by the real mechanism of 
the Sony Bank foreign exchange rate \cite{Sazuka,Sazuka2}. 
The Sony Bank rate is the foreign exchange rate 
that the Sony Bank offers with reference to the market rate. 
Basically, trades can be made on the web 
\cite{SonyBank} while the market is open. 
The Sony Bank rate 
depends on 
the market rate 
but is independent of 
the customers' orders. 
If 
the USD/JPY market rate 
changes by $\epsilon= \pm 0.1$ yen or more, 
the Sony Bank rate for 
USD/JPY is updated to the market rate. 
For instance, 
for the stochastic 
process of the real market 
(what we call {\it tick-by-tick data}): 
$X_{0}, 
X_{1}, X_{2}, 
\cdots,X_{T}$ with 
$|X_{1}|<\epsilon$ and 
$|X_{2}|>\epsilon$, 
the Sony Bank rate 
stays flat from 
the time 
$k=0$ to $k=1$, 
as the market rate 
is in the range of 
$\epsilon = \pm 0.1$ yen 
based on the market rate at 
$k=0$. 
When the market 
rate exceeds the range 
of $\epsilon=0.1$ yen at 
$k=2$, 
the Sony Bank rate 
is updated to the market rate. 
Obviously, 
the time interval 
$t=2-0=2$ here corresponds 
to the FPT we explained above and 
it is worth while for us to 
evaluate its distribution 
$P(t)$ to investigate statistical property of 
the Sony Bank USD/JPY rates. 

To calculate 
the FPT distribution 
$P(t)$ for 
the time 
series $X_{k}$, 
we define the 
probability 
$P(1)$, which 
means the probability of the FPT 
is $t=1$ as 
\begin{eqnarray}
P(1) & = & 
\lim_{T \to \infty}
\frac{1}{T}
\sum_{k=1}^{T}
\Theta 
(|X_{k+1}-X_{k}|-\epsilon)
\label{eq:P1}
\end{eqnarray}
where $\Theta(x)$ means 
the Heviside step function, namely, 
$\Theta (x)=1$ for 
$x \geq 0$ and $\Theta (x) =0$ for 
$x < 0$. 
We usually solve a kind of 
(fractal) Fokker-Plank 
equations 
under some appropriate 
boundary conditions 
\cite{Kappen,Tuckwell,Tuckwell2,Rangarajan,Rangarajan2,Rangarajan3} 
or use the so-called image method \cite{Redner,Durrett} 
to discuss 
the FPT problem. 
However, as we saw in 
equation (\ref{eq:P1}), 
our approach is completely 
different from 
such standard treatments. 
To evaluate 
the FPT (probability) distribution, 
say $P(1)$, 
we directly count 
the number of 
$t=1$, 
namely, 
${\cal N}_{1}=\sum_{k=1}^{T}\Theta(|X_{k+1}-X_{k}|-\epsilon)$
appearing within quite a long time interval $T$. 
We might choose $T$ 
as a time interval during which 
the market is open. 
Then, the ratio 
${\cal N}_{1}/T$ 
should be expected to converge to 
$P(1)$ as 
$T$ tends to 
infinity. 
This is the meaning of 
equation  
(\ref{eq:P1}) and is 
our basic idea for evaluating 
the FPT distribution. 
From our method, to 
evaluate the FPT distribution 
by counting ${\cal N}_{t}$ 
($t=1,2,\cdots$), the probability 
$P(2)$ is also given by 
$\lim_{T \to \infty}
({\cal N}_{2}/T)$, that is to say, 
\begin{eqnarray}
P(2) & = &  
\lim_{T \to \infty}
\frac{1}{T}
\sum_{k=1}^{T}
\Theta 
(|X_{k+2}-X_{k}|-\epsilon) - 
P(1) \nonumber \\
\mbox{} & = &  
\lim_{T \to \infty}
\frac{1}{T}
\sum_{k=1}^{T}
\Theta 
(|X_{k+2}-X_{k}|-\epsilon)
-
\lim_{T \to \infty}
\frac{1}{T}
\sum_{k=1}^{T}
\Theta 
(|X_{k+1}-X_{k}|-\epsilon).
\end{eqnarray}
In the same way as 
the probability 
$P(2)$, 
the probability 
$P(3)$ is obtained as 
\begin{eqnarray}
P(3) & = & 
\lim_{T \to \infty}
\frac{1}{T}
\sum_{k=1}^{T}
\Theta (
|X_{k+3}-X_{k}| - \epsilon) 
-P(1)-P(2) \nonumber \\
\mbox{} & = & 
\lim_{T \to \infty}
\frac{1}{T}
\sum_{k=1}^{T}
\Theta (
|X_{k+3}-X_{k}| - \epsilon) 
-
\lim_{T \to \infty}
\frac{1}{T}
\sum_{k=1}^{T}
\Theta (
|X_{k+2}-X_{k}| - \epsilon). 
\label{eq:P3}
\end{eqnarray}
We should notice that 
the probability $P(1)$ was cancelled 
in this expression (\ref{eq:P3}). 
Thus, we easily generalise this kind of 
calculations to 
evaluate the distribution $P(t)$ 
by repeating 
the above procedure 
as follows. 
\begin{eqnarray}
P(t) & = & 
\lim_{T \to \infty}
\frac{1}{T}
\sum_{k=1}^{T}
\Theta (
|X_{k+t}-X_{k}| - \epsilon) 
- 
\lim_{T \to 
\infty}
\frac{1}{T}
\sum_{k=1}^{T}
\Theta (
|X_{k+t-1}-X_{k}| - \epsilon) 
\label{eq:general_def}
\end{eqnarray}
where $P(1),\cdots,P(t-2)$ were all 
cancelled in this final formula 
(\ref{eq:general_def}). 
This equation 
(\ref{eq:general_def}) 
is the starting point of our evaluation. 
At a glance, 
this equation seems 
to be just a definition of 
the FPT distribution; 
however, 
for some classes of 
stochastic processes, 
we can derive 
the explicit form 
of 
the FPT distribution from 
this simple equation. 
In the next subsection, 
we derive 
the FPT distribution 
for the Wiener 
process as 
a simple test of 
our method. 
We stress that our approach helps as an intuitive account 
for the first passage process and 
derivation of its 
distribution. 
\subsection{A simple test of the method for Wiener stochastic processes}
To show the validity and usefulness 
of our method, we derive 
the FPT distribution 
from the above expression 
(\ref{eq:general_def}) for 
Wiener stochastic processes (Brownian motion). 
The ordinary Wiener 
process is described by 
$X_{t+1} = X_{t} + Y_{t}$, 
where the noise term 
$Y_{t}$ obeys the white Gaussian 
with zero-mean and variance $\sigma^{2}$. Then, 
we should notice that 
the difference 
$S_{t} \equiv 
X_{k+t}-X_{k}$ is rewritten 
in terms of the sum of the 
noise terms $Y_{t}$ as 
$S_{t} = \sum_{j=0}^{t-1}Y_{k+j}$. 
As is well-known, 
as the Gaussian process is stable, 
$S_{t}$ obeys 
the Gaussian with zero-mean and 
$\langle (S_{t})^{2} \rangle = 
t \sigma^{2}$ variance. Using the same 
argument as $S_{t}$, 
$S_{t-1}$ also obeys the Gaussian 
with 
zero-mean and 
$\langle (S_{t-1})^{2} \rangle = 
(t-1) \sigma^{2}$ variance. 
Therefore, the 
FPT distribution 
$P(t)$ derived by equation (\ref{eq:general_def}) leads to 
\begin{eqnarray}
P(t) & = & 
\Omega (p : S_{t},S_{t-1}) 
\label{eq:pt_integ} \\
\Omega(p : l,m) & \equiv & 
\int_{-\infty}^{\infty}
p(l) \Theta (|l|-\epsilon) dl 
-
\int_{-\infty}^{\infty}
p(m) 
\Theta (|m|-\epsilon) dm 
\label{eq:integ}
\end{eqnarray}
when we assume that the underlying 
stochastic process is ergodic, namely, 
that time average and ensemble average coincide. 
For the ordinary Wiener process, 
the probability 
distributions 
for 
$S_{t}$ and $S_{t-1}$ are 
Gaussians 
with zero mean and 
variances 
$t\sigma^{2},(t-1) \sigma^{2}$, 
respectively. Thus, we easily evaluate 
the integral appearing in 
(\ref{eq:integ}) 
after 
substituting 
$p=p_{G}(S_{t})=
(1/\sqrt{2\pi \sigma^{2}t})
\,{\rm e}^{-S_{t}^{2}/2\sigma^{2}t}, 
p_{G}(S_{t-1})=
(1/\sqrt{2\pi \sigma^{2} (t-1)})
\,{\rm e}^{-S_{t-1}^{2}/2\sigma^{2}(t-1)}$ 
and obtain 
$P(t)=\Omega(p_{G} : S_{t},S_{t-1})$ as 
\begin{eqnarray}
P(t) & = & 
2
\left\{
H
\left(
\frac{a}
{\sqrt{t}}
\right) -H 
\left(
\frac{a}
{\sqrt{t-1}}
\right)
\right\}
\label{eq:result_Wiener}
\end{eqnarray}
where 
we defined $a \equiv \epsilon/\sigma$. 
The function 
$H(x)$ is 
defined by 
$H(x) \equiv \int_{x}^{\infty}
dz\, 
{\rm e}^{-z^{2}/2}/\sqrt{2\pi}$. 
We should keep in 
mind that the above result is 
valid for discrete time 
$t$, however; 
it is easy for us to 
obtain its continuous 
time version by 
replacing 
$t \to t$, 
$t-1 \to t - \Delta t$ and evaluating  
(\ref{eq:result_Wiener}) 
in the limit of 
$\Delta t \to 0$. 
Then, we have 
\begin{eqnarray}
P(t) \Delta t & = & 
2H
\left(
\frac{a}
{\sqrt{t}}
\right) - 
2H 
\left(
\frac{a}
{\sqrt{t-\Delta t}}
\right) = 
2\frac{\partial}{\partial t}
H
\left(
\frac{a}
{\sqrt{t}}
\right) \Delta t 
+ 
{\cal O}((\Delta t)^{2})
\end{eqnarray}
Thus, 
the FPT distribution  
for the ordinary Wiener process in 
the continuous time limit is given by 
\begin{eqnarray}
P(t) & = & 
2\frac{\partial}{\partial t}
H
\left(
\frac{a}
{\sqrt{t}}
\right) = 
\frac{a\,{\exp}
\left(
-\frac{a^{2}}
{2t}
\right)}
{\sqrt{2\pi}\, t^{3/2}}
\label{eq:result_Wiener2}.
\end{eqnarray}
This well-known form 
is expected the inverse Gaussian 
distribution \cite{Chikara} 
for the FPT distribution 
of the ordinary Wiener process 
and 
is often observed 
in the so-called 
inter-spike 
interval (ISI) of 
the integrate-and-fire 
model for 
neural networks \cite{Tuckwell,Tuckwell2,Gerstner}. 
Therefore, 
in the asymptotic 
regime $t \to \infty$, 
the FPT distribution for the Wiener process 
obeys 
$t^{-3/2}$-scaling law. 
From the above discussion, 
we found that 
our new approach 
based on direct counting 
of the FPT 
to obtain the FPT distribution 
is effective and gives 
a well-known solution 
for the ordinary Wiener process. 

Before we move to the main section, 
we should comment on the much shorter 
derivation of the above formulation (\ref{eq:result_Wiener2}). 
Let us define our stochastic process by 
$x(t)$ and the probability density function 
finding the process in $x^{'}$ at time $t$ 
provided that it was in $x$ at time $0$ by 
$p(x^{'},t|x,0)$. 
Here, we assume that $p(x^{'},t|x,0)$ is symmetric 
with respect to $x^{'}=0$. 
We also define the FPT 
with absorbing barriers $x=\pm \epsilon$ by 
$\tau$. 
Then, the complementary cumulative 
distribution of $\tau$ is given by 
\begin{eqnarray}
W(x,t) & = & 
P_{r}(\tau \geq t) = 
\int_{-\epsilon}^{\epsilon}p(x^{'},t|x,0)dx^{'} 
\end{eqnarray}
and the cumulative distribution function leads to 
\begin{eqnarray}
H(x,t) & = & 
P_{r}(\tau <t)=
1-W(x,t)=2
\int_{\epsilon}^{\infty}p(x^{'},t|x,0)dx^{'}.
\end{eqnarray}
Thus, 
the FPT distribution is obtained by 
\begin{eqnarray}
p(x,t) & = & 
\frac{\partial H(x,t)}{\partial t}.
\end{eqnarray}
This equation coincides with 
equation (\ref{eq:result_Wiener2}). 
We should also 
mentioned that 
the similar derivations are found in 
the references \cite{Risken,Gardiner}. 
\section{Stable processes and their FPT distributions}
We stress that 
our method is widely applicable 
to stochastic processes 
whose distributions are stable. 
Stable processes are 
specified as follows. 
If stochastic variables 
$Y_{i}$ ($i=1,\cdots,N$) are 
identically independently 
distributed from 
$p(Y_{i})$, 
the Fourier 
transform of 
the sum of 
the $Y_{i}$, 
namely, 
$S_{n}=\sum_{j=1}^{N}
Y_{j}$ is 
given by 
\begin{eqnarray}
\varphi_{n}(q) & = & 
[\varphi (q)]^{n}
\label{eq:irv}
\end{eqnarray}
where 
$\varphi(q)$ is the Fourier transform of 
the stochastic variable $Y$, 
namely, 
the characteristic function and  
defined by 
\begin{eqnarray}
\varphi (q) & = &  
\int_{-\infty}^{\infty}
p(Y)\, 
{\rm e}^{iqY}
dY. 
\end{eqnarray}
Then, the stochastic 
process 
$Y_{i}$ is referred to as a {\it stable process}. 
Strictly speaking, 
equation (\ref{eq:irv}) is a 
possible definition of 
infinitely divisible random variables and 
not of stable random variables. 
A stable random variable is infinitely 
divisible and stability refers to 
the invariance of the distribution  
with respect to convolutions. 

It is obvious that 
for this class of stable processes, 
the FPT 
distribution 
is easily obtained by 
our method 
because 
the probability distributions 
$p(S_{t})$ and 
$p(S_{t-1})$ to 
evaluate 
$\Omega (p : S_{t},S_{t-1})$ 
in (\ref{eq:pt_integ}) are defined explicitly. 
In the next subsections, 
we show several results 
from our new approach. 
\subsection{Lorentzian stochastic processes}
As a first simple example 
of the stable distributions, 
let us think about 
Lorentzian stochastic processes: 
$X_{t+1}=X_{t}+Y_{t}$, 
where 
the noise term 
$Y_{t}$ 
obeys the 
following 
white 
Lorentzian :
\begin{eqnarray}
p(Y_{t}) & = & 
\frac{\gamma}{\pi}
\frac{1}{\gamma^{2}+Y_{t}^{2}}.
\end{eqnarray}
Then, the characteristic function 
of the stochastic variable 
$Y_{t}$ is given by 
\begin{eqnarray}
\varphi (q) & = & 
\frac{\gamma}{\pi}
\int_{-\infty}^{\infty}
\frac{{\rm e}^{iqY_{t}}}
{\gamma^{2}+Y_{t}^{2}}\, 
dY_{t} = 
{\rm e}^{-\gamma|q|}. 
\end{eqnarray}
By using 
the convolution 
of the Fourier 
transform for 
the variable 
$S_{n}=\sum_{k=1}^{n}Y_{k}$, 
we have $\varphi_{n}(q)=[\varphi (q)]^{n}=
{\rm e}^{-n \gamma |q|}$. 
Therefore, 
the inverse Fourier transform 
of $\varphi_{n}(q)$ leads to 
the 
probability distribution 
of the sum of noise term as follows.  
\begin{eqnarray}
p_{Lorentz}(S_{n}) & = & 
\frac{1}{2\pi}
\int_{-\infty}^{\infty}
{\rm e}^{-n \gamma |q|-iqx}
dq = 
\frac{\gamma n}{\pi}
\frac{1}
{(\gamma n)^{2}+S_{n}^{2}}
\end{eqnarray}
By substituting 
this probability 
$p_{Lorentz}(S_{t}), 
p_{Lorentz}(S_{t-1})$ into 
equations 
(\ref{eq:pt_integ}) and (\ref{eq:integ}), 
we obtain the 
FPT distribution 
for the Lorentzian stochastic processes 
as follows: 
\begin{eqnarray}
P(t) & = & 
\Omega (p_{Lorentz} : S_{t},S_{t-1}) = 
2\int_{\epsilon}^{\infty}
\frac{\gamma t}
{\pi}
\frac{dS_{t}}
{(\gamma t)^{2}+S_{t}^{2}} - 
2\int_{\epsilon}^{\infty}
\frac{\gamma (t-1)}
{\pi}
\frac{dS_{t-1}}
{(\gamma (t-1))^{2} + 
S_{t-1}^{2}} \nonumber \\
\mbox{} & = & 
\frac{2}{\pi}
\tan^{-1}
\left(
\frac{b}{t-1}
\right) - 
\frac{2}{\pi}
\tan^{-1}
\left(
\frac{b}{t}
\right)
\end{eqnarray}
where 
we defined $b \equiv 
\epsilon/\gamma$. 
This is a result for discrete time 
steps; however, 
its continuous version is 
easily obtained 
by using 
$d(\tan^{-1}(y))/dy = 
(1+y^{2})^{-1}$ as 
follows. 
\begin{eqnarray}
P(t) & = & 
-\frac{\partial}
{\partial t}
\left\{
\frac{2}{\pi}
\tan^{-1}
\left(
\frac{b}{t}
\right)
\right\} = 
\frac{2b}{\pi}
\frac{1}{b^{2}+t^{2}}
\label{eq:pt_Lorentz}
\end{eqnarray}
From this 
result (\ref{eq:pt_Lorentz}), 
we find that 
the FPT distribution for 
the Lorentzian stochastic 
processes obeys 
Lorentzian.  
In the asymptotic regime 
$t \to \infty$, 
the FPT distribution 
for the Lorentzian 
stochastic processes  
obeys the $t^{-2}$-scaling law. 
\subsection{Anomalous diffusion of L$\acute{\rm e}$vy flight}
We next consider the case of 
L$\acute{\rm e}$vy stochastic 
processes 
whose 
noise term $Y_{t}$ of the 
stochastic 
process $X_{t+1}=X_{t}+Y_{t}$ 
obeys 
the following L$\acute{\rm e}$vy distribution : 
\begin{eqnarray}
p_{L\acute{\rm e}vy}(Y_{t}) & = & 
\frac{1}{\pi}
\int_{0}^{\infty}
{\rm e}^{-\gamma |q|^{\alpha}}
\cos 
(q Y_{t})\,dq 
\label{eq:LEVY}
\end{eqnarray}
We should keep in mind that 
the above 
distribution (\ref{eq:LEVY}) 
is reduced to  
the Wiener stochastic process 
($\alpha=2$) and 
the Lorentzian stochastic 
process ($\alpha=1$) as 
its special cases. 
As this process 
$Y_{t}$ ($t=1,\cdots,n$) 
is also stable, 
the sum of 
the noise term 
$S_{n} = \sum_{j=1}^{n}
Y_{j}$ 
has the following 
probability 
distribution 
\begin{eqnarray}
p_{L\acute{\it e}vy}(S_{n}) & = & 
\frac{1}{\pi}
\int_{0}^{\infty}
{\rm e}^{-\gamma n |q|^{\alpha}}
\cos 
(q S_{n})\,dq.
\end{eqnarray}
Now, we can derive the 
FPT distribution 
by substituting 
$p_{L\acute{\it e}vy}(S_{t}), 
p_{L\acute{\it e}vy}(S_{t-1})$ into 
equations 
(\ref{eq:pt_integ}) and (\ref{eq:integ}) as 
\begin{eqnarray}
P(t) & = & 
\Omega(p_{L\acute{\it e}vy},S_{t},S_{t-1}) \nonumber \\
\mbox{} & = & 
\frac{2}{\pi}
\int_{\epsilon}^{\infty}
dS 
\int_{0}^{\infty}
{\rm e}^{-\gamma t |q|^{\alpha}}
\cos 
(q S)\,dq - 
\frac{2}{\pi}
\int_{\epsilon}^{\infty}
dS 
\int_{0}^{\infty}
{\rm e}^{-\gamma (t-1) |q|^{\alpha}}
\cos (q S) \,dq
\label{eq:Pt_LEVY2}. 
\end{eqnarray}
Expression of 
a continuous 
time version 
(\ref{eq:Pt_LEVY2}) 
is obtained 
from 
the derivative of 
the above discrete time 
distribution $P(t)$ with respect to 
$t$ as 
\begin{equation}
P(t) =  
\frac{\partial}
{\partial t}
\left\{
\frac{2}{\pi}
\int_{\epsilon}^{\infty}
dS 
\int_{0}^{\infty}
{\rm e}^{-\gamma t |q|^{\alpha}}
\cos 
(q S) \,dq
\right\} = 
-\frac{2\gamma}{\pi}
\int_{\epsilon}^{\infty}
dS 
\int_{0}^{\infty}
|q|^{\alpha}
\,
{\rm e}^{-\gamma t |q|^{\alpha}}
\cos (q S) \, dq.
\end{equation}
In the asymptotic 
regime $t \to \infty$, 
by replacing the variable 
as $tq^{\alpha}=Q$ and 
after some simple algebra, 
we obtain 
\begin{eqnarray}
P(t) & = & 
\frac{2\gamma t^{-(\alpha + 1)/\alpha}}{\pi \alpha}
\int_{\epsilon}^{\infty}
dS 
\int_{0}^{\infty}
Q^{1/\alpha}
{\rm e}^{-\gamma Q}
\cos
\left[
\pi 
+
\left(
\frac{Q}{t}
\right)^{1/\alpha}
S
\right]dQ \nonumber \\
\mbox{} & = & 
\Psi (\alpha,\gamma)\,
t^{-(\alpha +1)/\alpha} 
\label{eq:pt_Levy} \\
\Psi (\alpha,\gamma)  & \equiv & 
\frac{2\gamma}{\pi \alpha^{2}}
\sum_{l=0}^{\infty}
\gamma^{-(2l+1)/\alpha}
\frac{(-1)^{l+1}
\rho^{2l+1}}
{(2l)!}
\Gamma
\left(
\frac{2l + 1}
{\alpha}
\right)
\label{eq:pt_Levy2}
\end{eqnarray}
where the $\Gamma(x)$ means 
Gamma function 
and $\rho$ is a constant of order $1$. 
Then, 
we should notice 
that the 
above scaling law 
is 
consistent with 
both 
the Wiener 
stochastic 
process 
($P(t) \sim 
t^{-3/2}$ for $\alpha=2$) and with  
the Lorentzian stochastic 
process ($P(t) \sim t^{-2}$ for $\alpha=1$) 
we discussed in the previous subsections. 

As we saw in the above stochastic 
processes, 
our new approach 
based on direct counting of 
the FPT is widely useful for the class of 
stable stochastic processes and 
the final expressions 
(\ref{eq:pt_integ})(\ref{eq:integ}), 
containing at most just two integrals. 
Moreover, 
our approach can be applied to 
the FPT problems with a surprisingly 
wide variety of absorbing 
boundary conditions. 
This is one of the advantages of 
our method over other approaches based on 
analysis of Fokker-Plank equations.  
To show the advantage, in the next section, 
we apply our method to much more complicated 
stochastic stable process. 
\subsection{Computer simulations}
\begin{figure}[ht]
\begin{center}
\rotatebox{-90}{\includegraphics[width=5.65cm]{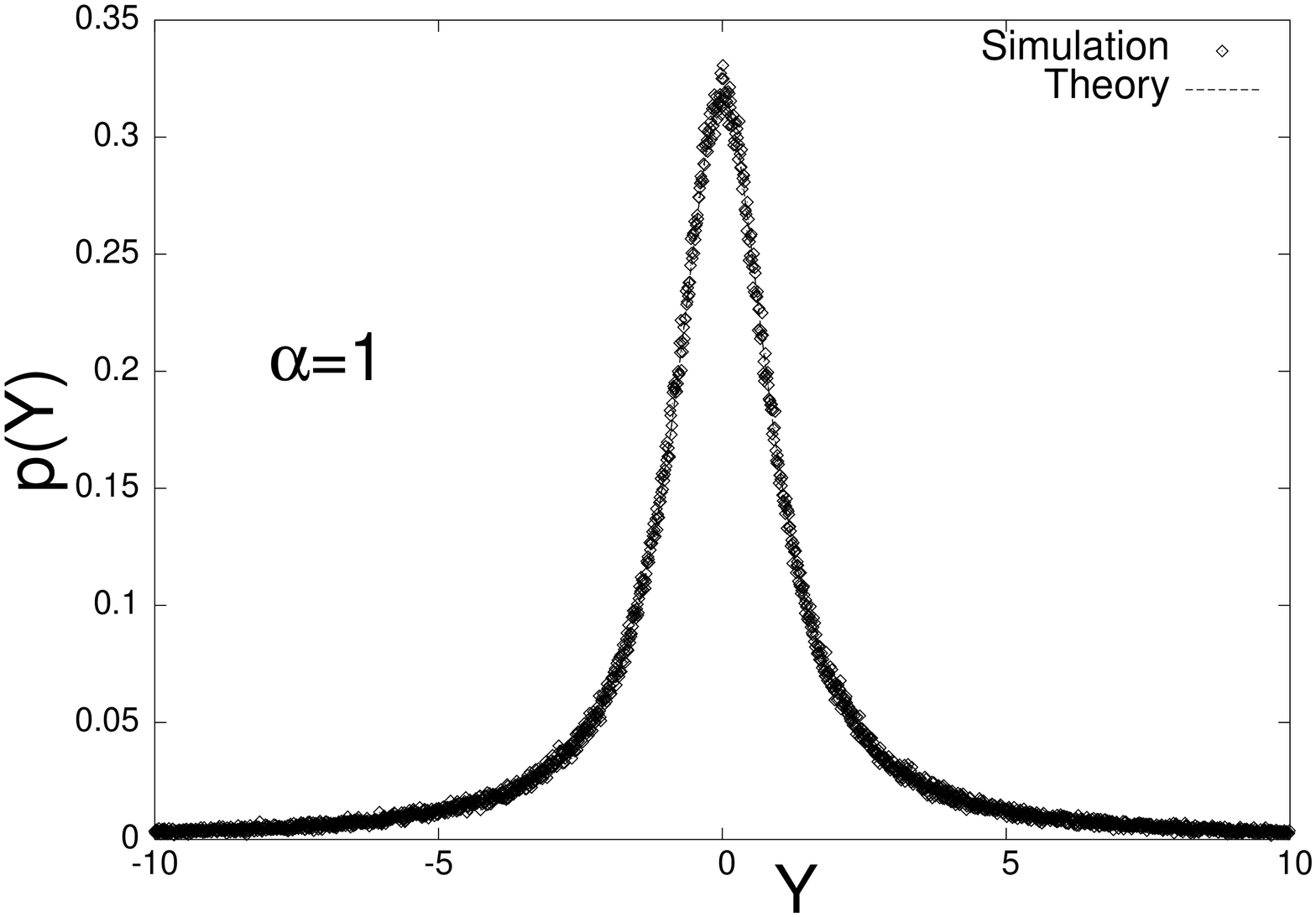}}
\rotatebox{-90}{\includegraphics[width=5.65cm]{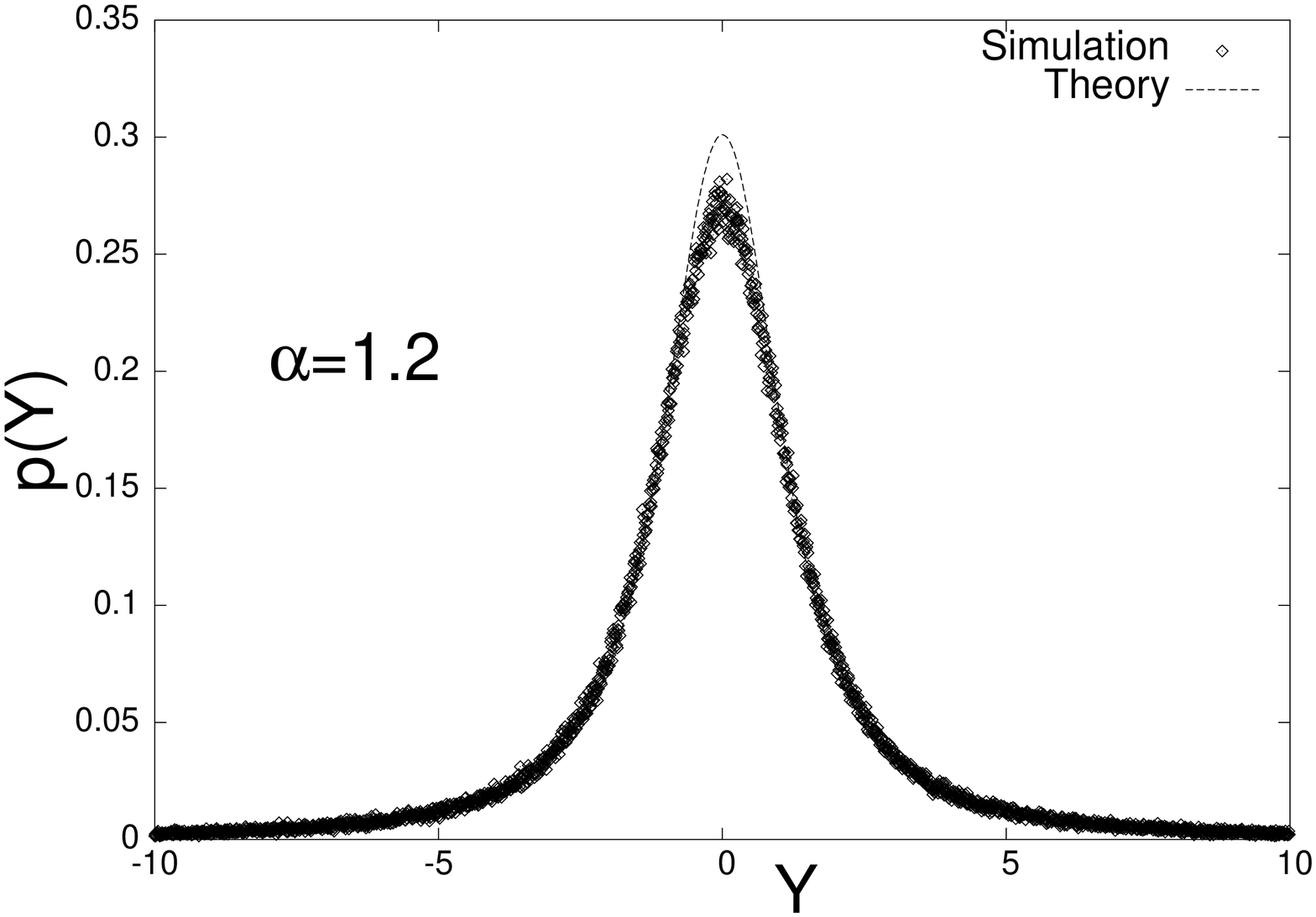}}
\end{center}
\caption{\footnotesize 
L$\acute{\rm e}$vy distributions 
obtained by the chaotic map 
(\ref{eq:chaos_map}) with 
$\alpha=1$ (Lorentzian: left) and 
$\alpha=1.2$ (right). 
The broken lines are 
corresponding theoretical results 
given by 
(\ref{eq:theor_Levy}). 
We set the number of the superposition $N=10000$. 
}
\label{fig:fg_add1}
\end{figure}

\begin{figure}[ht]
\begin{center}
\rotatebox{-90}{\includegraphics[width=5.65cm]{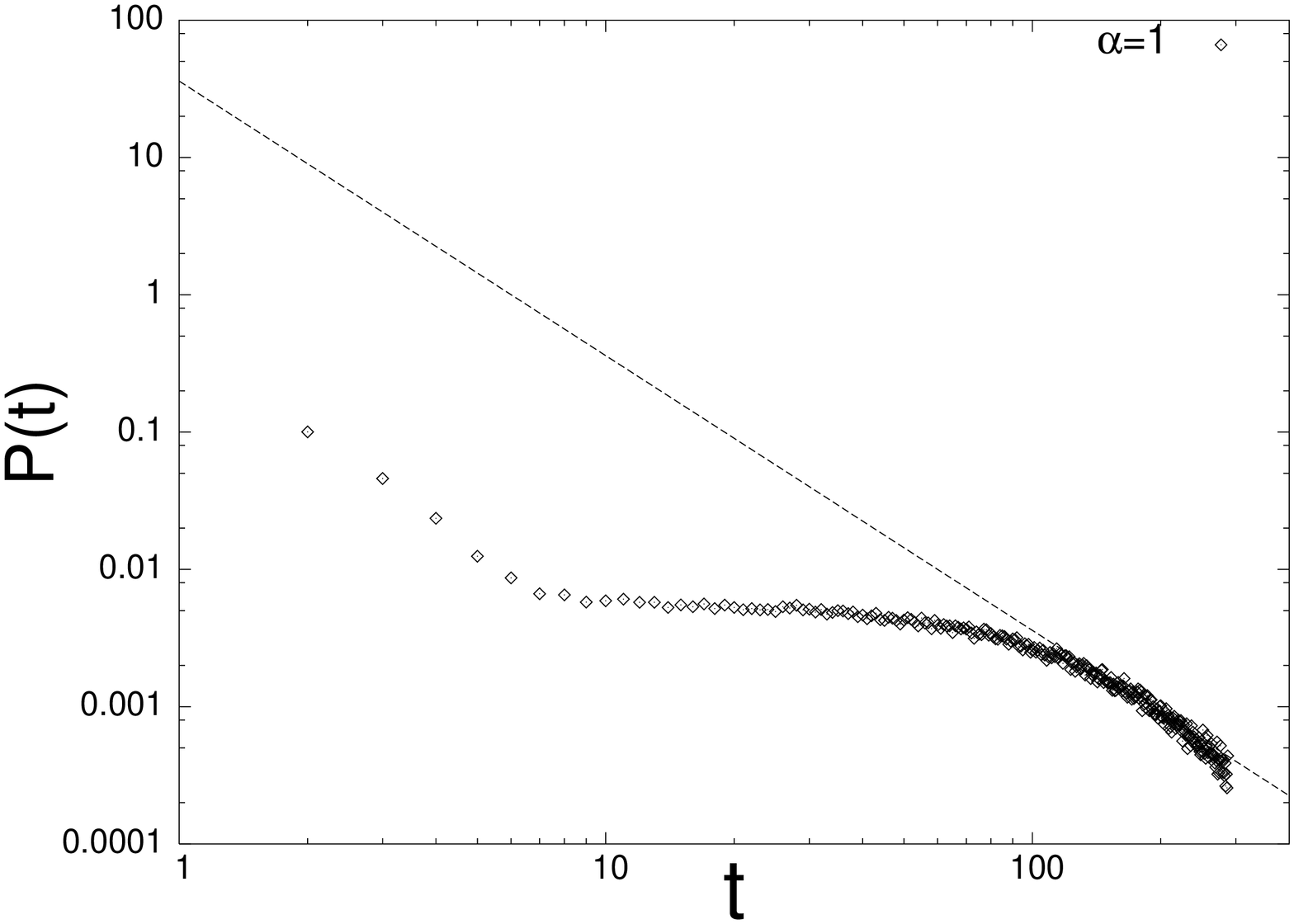}}
\rotatebox{-90}{\includegraphics[width=5.65cm]{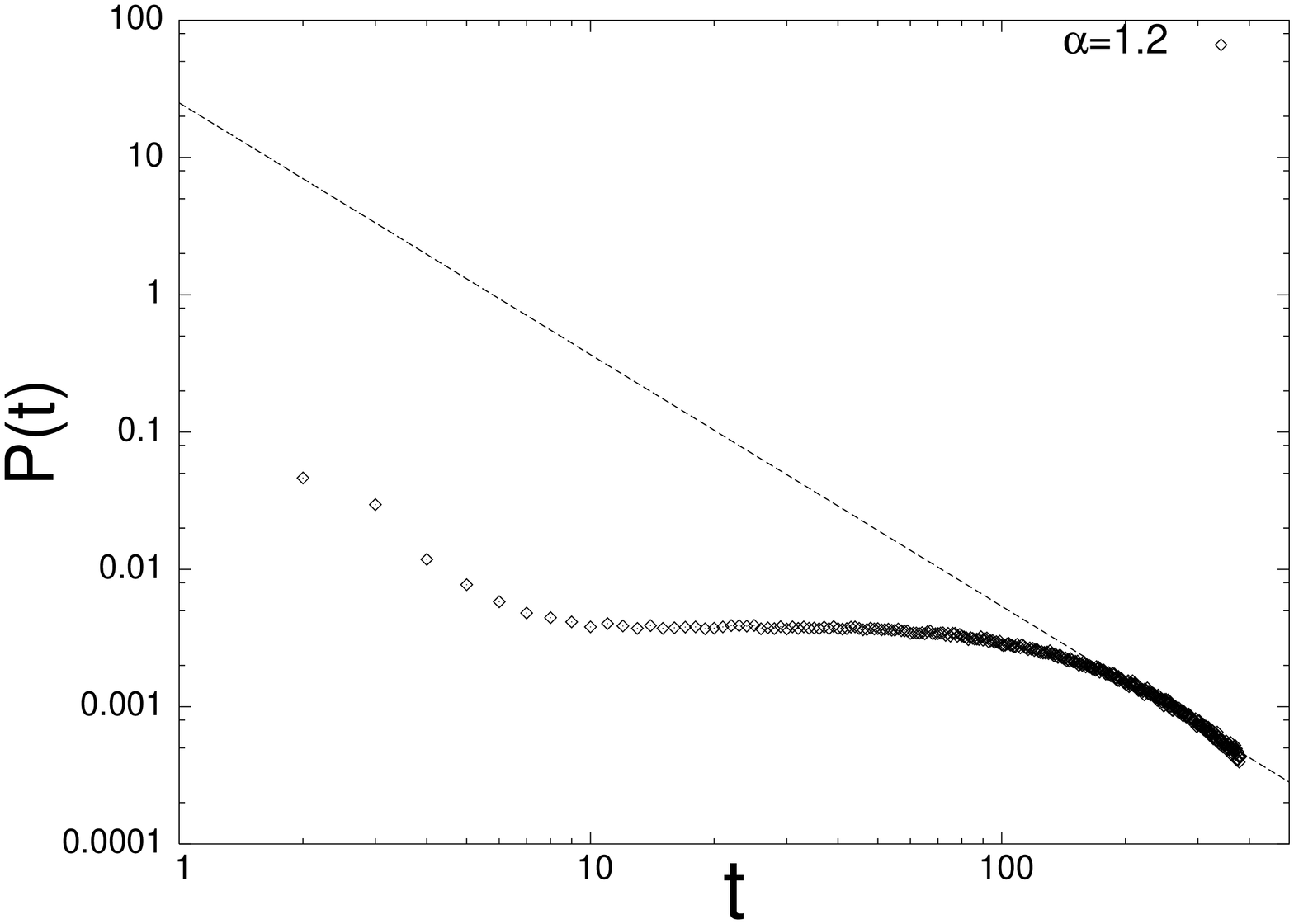}}
\end{center}
\caption{\footnotesize 
The tail exponents of 
the FPT distributions for 
$\alpha=1$ (left) and 
$\alpha=1.2$ (right). 
The values of the power-law 
exponent predicted by our theory 
(the slope of the broken line in each panel) 
are $-2$ for $\alpha=1$ and $-1.8333$ for 
$\alpha=1.2$, respectively. 
We set the width of the rate window $\epsilon=120$. 
}
\label{fig:fg_add2}
\end{figure}

To check the theoretical prediction 
of the power-law exponent for the 
FPT distribution, we perform computer simulations. 
Then, to generate the additive noise 
$Y_{t}$ for each time step, the L$\acute{\rm e}$vy 
stable distribution $p(Y)$ is needed. 
As shown by Umeno \cite{Umeno}, 
the L$\acute{\rm e}$vy stable distribution 
is obtained by a superposition of a chaotic map such as 
\begin{eqnarray}
Y_{n+1} & = & 
\left|
\frac{1}{2}
\left(
|Y_{n}|^{\alpha}
-\frac{1}{|Y_{n}|^{\alpha}}
\right)
\right|^{1/\alpha}
{\rm sgn}
\left(
Y_{n}-\frac{1}{Y_{n}}
\right)\,\,\,\,\,\,\,\,\,\,\,
(0 < \alpha < 2)
\label{eq:chaos_map}, 
\end{eqnarray}
namely, 
for independently selected initial conditions 
$i=1,\cdots,N$, 
the quantity 
$\sum_{i=1}^{N}Y^{(i)}/N^{1/\alpha}$ 
obeys a L$\acute{\rm e}$vy stable law with 
$\alpha$, that is, 
\begin{eqnarray}
p(Y) & = & 
\frac{1}{\pi}
\int_{0}^{\infty}
{\rm e}^{-q^{\alpha}}
\cos (qY)dq
\label{eq:theor_Levy}.
\end{eqnarray}
In FIG. \ref{fig:fg_add1}, 
we plot the 
L$\acute{\rm e}$vy 
distribution with $\alpha=1$ (Lorentzian) 
and $\alpha=1.2$ obtained by the 
superposition of the chaotic map 
(\ref{eq:chaos_map}). 
We set the number of the superposition 
of the chaotic map $N=10000$.
From this figure, we find that 
the distributions obtained by the simulations 
are in good agreement with the corresponding 
analytical expressions (\ref{eq:theor_Levy}). 
Keeping these results in mind, 
we use the sampling point from 
the superposition of the chaotic map 
(\ref{eq:chaos_map}) as the additive noise 
$Y_{t}$ in the stochastic process for each time step. 
Then, we should notice that 
one should choose a large value for 
the width of the rate window $\epsilon$, say, 
$\epsilon \sim 100$ to 
investigate the tail of the FPT distribution. 
This is because it is a rare event 
for the random walker to stay in the range $[-\epsilon,\epsilon]$ 
and it takes quite long computational time for us 
to obtain the tail exponent (relatively long 
FPT) if $\epsilon$ is small. 
Here, we choose $\epsilon=120$ and 
evaluate the power-law exponent of the FPT distribution 
for the L$\acute{\rm e}$vy processes 
with $\alpha=1$ and $\alpha=1.2$. 
The results are shown in 
FIG. \ref{fig:fg_add2}. 
From this figure, 
we find that the power-law tails have almost 
the same exponents as those predicted 
by our theory. 
Of course, the tail region is too noisy to conclude 
that the simulations are completely consistent with 
the theory. However, at least, we may say that 
they are not in disagreement. 
Thus, the computer simulations provided us 
with a justification of our theoretical formulation. 
\section{Crossover in scaling laws of FPT distributions}
In the previous section, 
we showed our new formulation 
is effective and much more simpler than the 
approach of the (fractal) Fokker-Plank 
equations \cite{Rangarajan} 
to obtain the 
FPT distribution for 
stable stochastic processes. 
We actually found that 
the FPT distribution 
of 
the general non-Gaussian 
L$\acute{\rm e}$vy 
stochastic process 
specified by parameter 
$\alpha$ is 
obtained and its 
scaling behaviour 
in the 
asymptotic 
regime $t \to \infty$ is 
$t^{-(\alpha +1)/\alpha}$-law. 
In this section, 
we show that 
our formalism is 
also 
useful in obtaining the FPT distribution 
for the so-called {\it truncated 
L$\acute{\rm e}$vy flight} 
(the so-called KoBoL processes from 
Koponen, Boyarchenko and Levendorskii \cite{Koponen,KoBoL,Boyarchenko}), 
for which it is well-known 
that the crossover 
between the 
a L$\acute{\rm e}$vy 
and a Gaussian regime 
in the 
distribution 
of the real time step 
takes place \cite{Mantegna94,Mantegna94_2,Koponen,Mantegna2000}. 
In this section, 
we show, by using 
our method based on 
direct counting of the FPT, 
this kind of crossover 
in scaling laws 
is also observed 
in the FPT $t$.   

The characteristic function 
for the truncated 
L$\acute{\rm e}$vy flight 
is 
defined by 
\begin{eqnarray}
T(q) & = & 
{\exp}
\left[
-\gamma 
\frac{(\Delta^{2}+|q|^{2})^{\alpha/2}
\cos 
\left(
\alpha 
\tan^{-1}
\left(
\frac{|q|}
{\Delta}
\right)
\right)-
\Delta^{\alpha}}
{
\cos
\left(
\frac{\pi \alpha}{2}
\right)
}
\right]
\label{eq:TLF}
\end{eqnarray}
where $\tan^{-1}(|q|/\Delta) \simeq 
\pi/2$ as 
$\Delta \to 0$ \cite{Voit}. 
Therefore, 
in this limit 
$\Delta \to 0$, the above equation 
(\ref{eq:TLF}) 
is reduced to $T(q)={\rm e}^{-\gamma |q|^{\alpha}}$. 
Obviously, 
this is identical to 
the characteristic 
function of the conventional 
L$\acute{\rm e}$vy 
flight as we already 
saw in the previous section. 
For this reason, 
a non-zero 
value of 
$\Delta$ 
controls the cut-off width of the 
truncated 
L$\acute{\rm e}$vy 
flight. 
We should notice that one could 
also use 
a hard cut-off version 
of the truncation scheme \cite{Mantegna94}, 
namely, 
\begin{eqnarray}
p_{TLF}(Y_{t}) & = & 
p_{L\acute{\rm e}vy}(Y_{t}) 
\Theta (\Delta^{-1}-|Y_{t}|).
\end{eqnarray}
However, 
for its mathematical simplicity, 
we use the soft cut-off version 
of the truncation scheme which will be 
explained below. 

By the convolution 
of the Fourier 
transform, we can show 
\begin{eqnarray}
T_{n}(q) & = & 
[T(q)]^{n} = 
{\exp}
\left[
-n \gamma 
\frac{(\Delta^{2}+|q|^{2})^{\alpha/2}
\cos 
\left(
\alpha 
\tan^{-1}
\left(
\frac{|q|}
{\Delta}
\right)
\right)-
\Delta^{\alpha}}
{
\cos
\left(
\frac{\pi \alpha}{2}
\right)
}
\right]
\end{eqnarray}
and then the sum of 
the noise 
term 
$S_{n}=\sum_{j=1}^{n}Y_{j}$ 
of the 
truncated L$\acute{\rm e}$vy flight obeys 
the following 
probability 
distribution: 
\begin{eqnarray}
p_{TLF}(S_{n}) & = & 
\frac{1}{2\pi}
\int_{-\infty}^{\infty}
dq \, 
{\rm e}^{-i q S_{n}}
[T(q)]^{n} \nonumber \\
\mbox{} & = & 
\frac{1}{\pi}
\int_{0}^{\infty}
dq\,
\cos (qS_{n}) 
\left[
-n \gamma 
\frac{(\Delta^{2}+|q|^{2})^{\alpha/2}
\cos 
\left(
\alpha 
\tan^{-1}
\left(
\frac{|q|}
{\Delta}
\right)
\right)-
\Delta^{\alpha}}
{
\cos
\left(
\frac{\pi \alpha}{2}
\right)
}
\right]
\end{eqnarray}
Substituting these
probability 
distributions $p_{TLF}(S_{t}), 
p_{TLF}(S_{t-1})$ into 
equations 
(\ref{eq:integ}) 
and (\ref{eq:pt_integ}), 
and taking the derivative 
of $P(t)$ with respect to 
$t$, we obtain 
the FPT distribution 
of the truncated L$\acute{\rm e}$vy 
flight for continuous time case as 
\begin{eqnarray}
P(t) & = & 
-\frac{2\gamma}{\pi}
\int_{\epsilon}^{\infty}dS 
\int_{0}^{\infty}
dq 
\left[ 
\frac{(\Delta^{2}+|q|^{2})^{\alpha/2}
\cos 
\left(
\alpha 
\tan^{-1}
\left(
\frac{|q|}
{\Delta}
\right)
\right)-
\Delta^{\alpha}}
{
\cos
\left(
\frac{\pi \alpha}{2}
\right)
}
\right] \nonumber \\
\mbox{} & \times & 
{\exp}
\left[
-\gamma t 
\frac{(\Delta^{2}+|q|^{2})^{\alpha/2}
\cos 
\left(
\alpha 
\tan^{-1}
\left(
\frac{|q|}
{\Delta}
\right)
\right)-
\Delta^{\alpha}}
{
\cos
\left(
\frac{\pi \alpha}{2}
\right)
}
\right] \cos (qS).
\label{eq:pt_TLF}
\end{eqnarray}
Thus far, it has been difficult to perform 
the above two 
integrals with 
respect to 
$S$ and $q$ analytically to 
obtain a compact form of 
the FPT distribution. 
However, numerical integrations 
of equation (\ref{eq:pt_TLF}) 
enable us to proceed to it.  
\begin{figure}[ht]
\begin{center}
\rotatebox{-90}{\includegraphics[width=5.65cm]{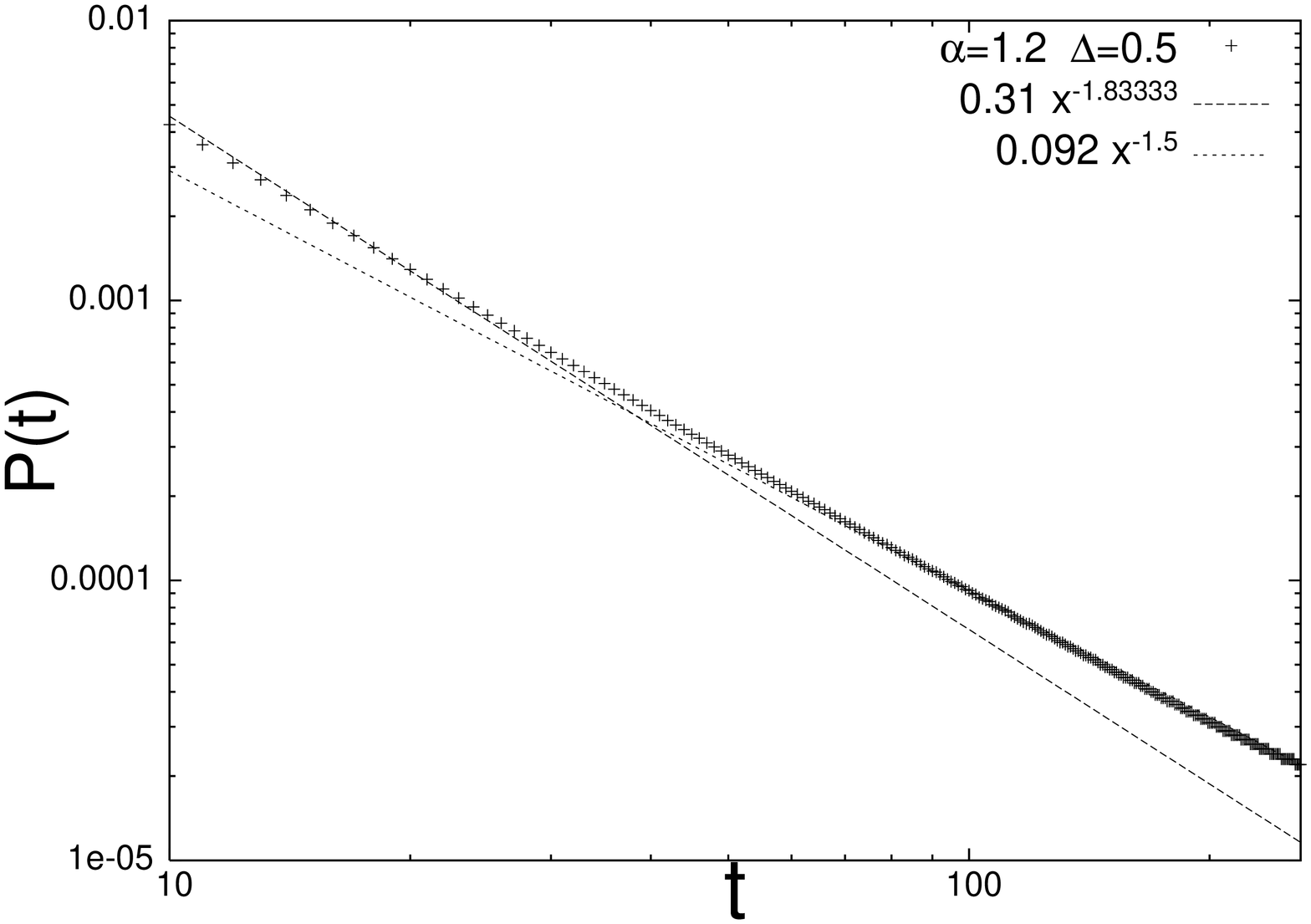}}
\rotatebox{-90}{\includegraphics[width=5.65cm]{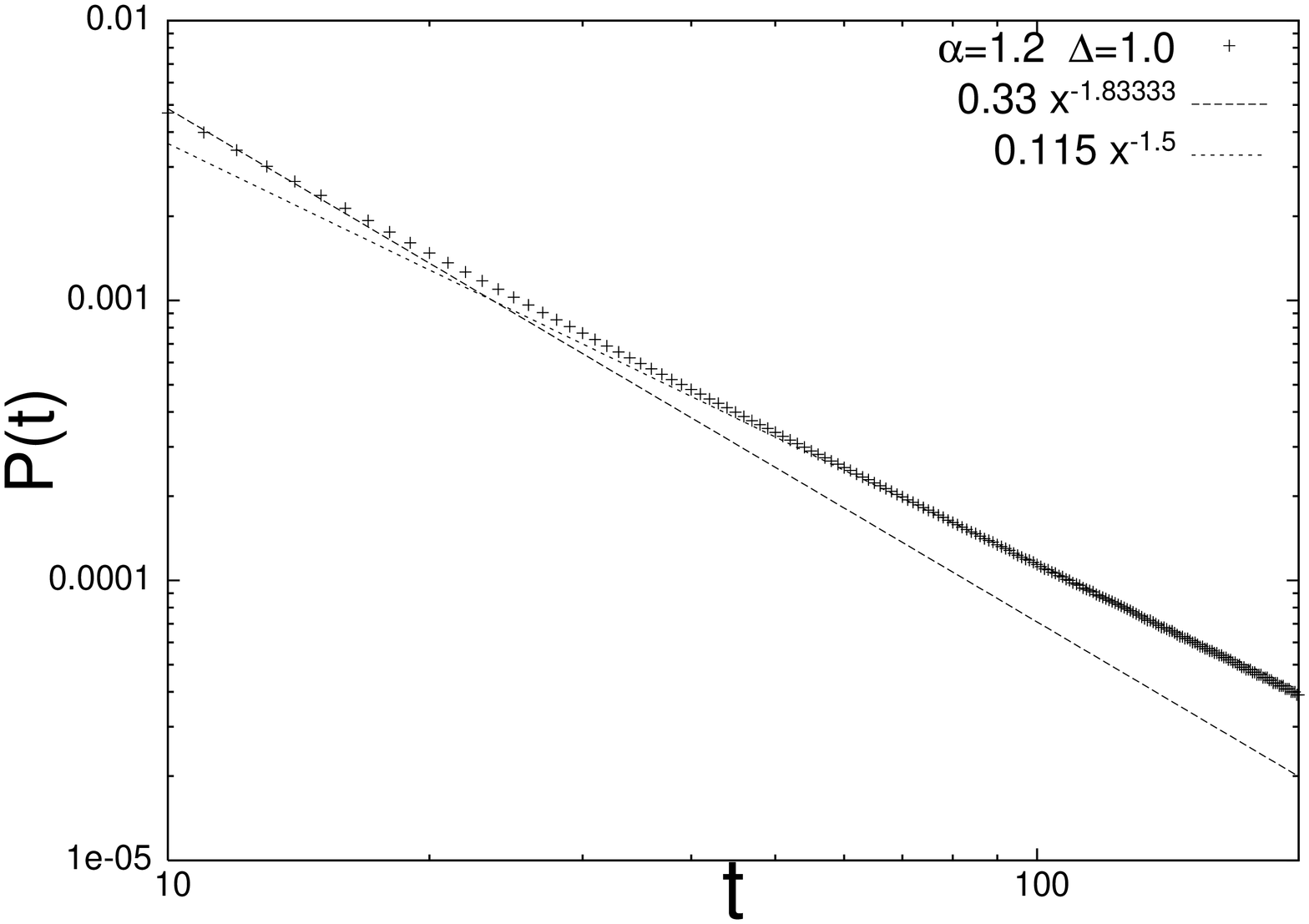}}
\rotatebox{-90}{\includegraphics[width=5.65cm]{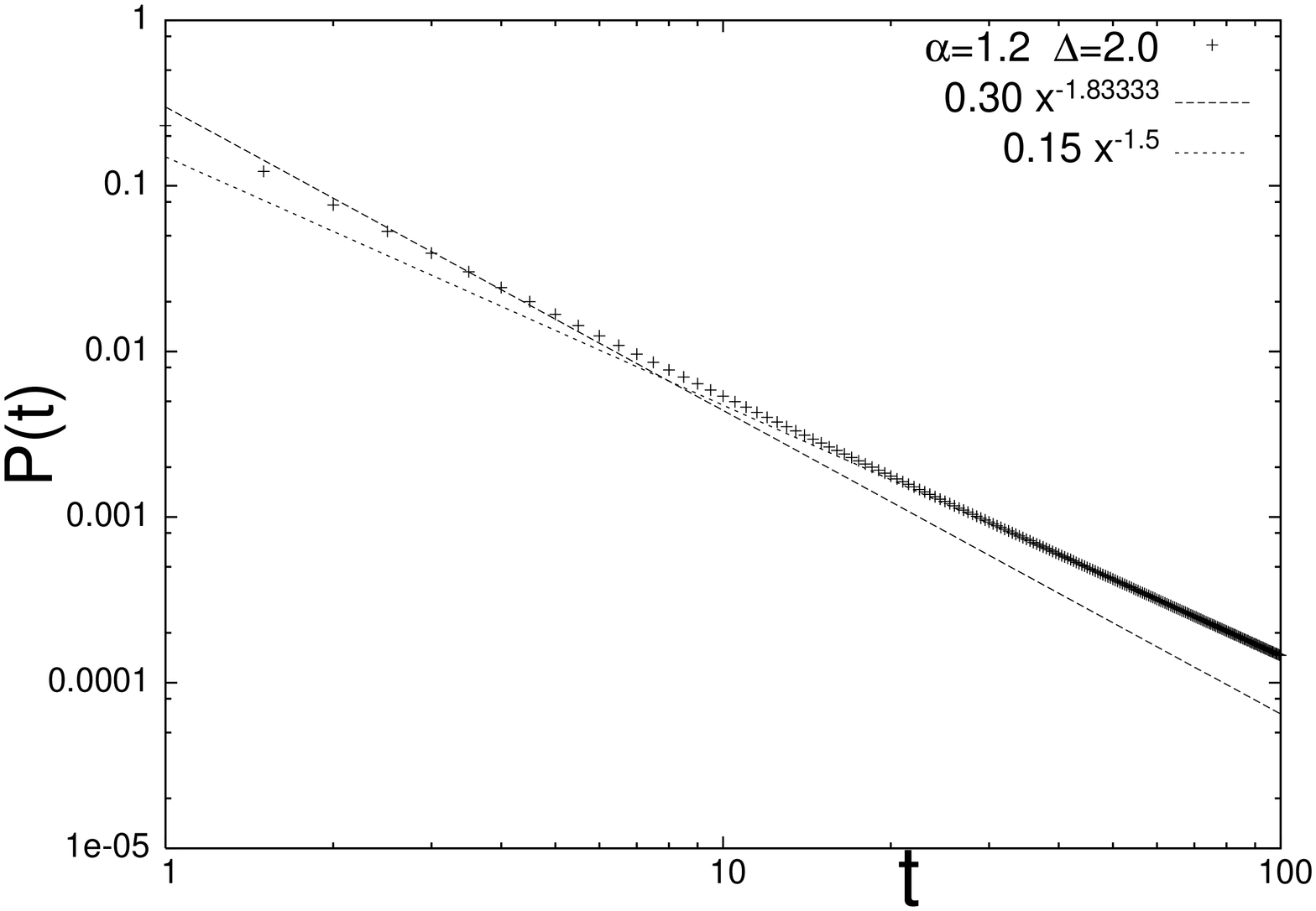}}
\end{center}
\caption{\footnotesize 
Scaling laws 
of the FPT distribution 
for the truncated L$\acute{\rm e}$vy flight.
We set 
$\alpha=1.2$ and 
$\Delta=0.5$ (upper left panel), 
$1.0$ (upper right panel) and 
$\Delta=2.0$ (lower panel). 
We find a clear 
crossover between L$\acute{\rm e}$vy 
and Gaussian regimes. 
The scaling laws change 
at the 
crossover point 
$t_{\times} \simeq 
44$ ($\Delta=0.5$), 
$19$ ($\Delta=1.0$) and 
$t_{\times} \simeq 8$ ($\Delta=2.0$). 
Below 
the crossover point 
$t_{\times}$, 
the scaling laws 
are 
those of 
the L$\acute{\rm e}$vy : 
$t^{-(\alpha +1)/\alpha}=t^{-1.83}$, whereas, 
above 
$t_{\times}$, 
the scaling laws 
become 
those 
of the 
Gaussian : $t^{-3/2}$.
In these three 
panels, 
we find that the scaling relation : 
$t_{\times}(\Delta=0.5)/t_{\times}(\Delta=1.0)=
t_{\times}(\Delta=1.0)/t_{\times}(\Delta=2.0)= 
2^{\alpha} \simeq 2.297$ holds 
for the crossover points. 
}
\label{fig:fg1}
\end{figure}
In FIG. \ref{fig:fg1}, 
we show the 
scaling plot of the 
FPT distribution for 
the truncated L$\acute{\rm e}$vy flight 
with $\alpha=1.2$ for 
several values of 
$\Delta$. 
From these three 
panels in FIG. \ref{fig:fg1}, 
we find that 
the scaling law of 
the FPT distribution 
changes 
from 
$t^{-(\alpha +1)/\alpha}=
t^{-(1.2+1)/1.2} \sim 
t^{-1.83}$ to 
$t^{-3/2}$ at some crossover points $t_{\times}
\simeq 44$ ($\Delta=0.5$), 
19 ($\Delta=1.0$) and 
$t_{\times} \simeq 
8$ ($\Delta=2.0$). 
To obtain 
useful information 
about 
the crossover point 
$t_{\times}$, 
we evaluate 
the asymptotic 
form of 
the FPT distribution 
(\ref{eq:pt_TLF}) for 
both 
$|q| \leq \Delta$ (Gaussian 
regime) and 
$|q| > \Delta$ (non-Gaussian 
L$\acute{\rm e}$vy regime). 

For Gaussian 
regime $|q| \leq \Delta$, 
by replacing 
the variable 
$q$ with 
$Q$ as 
$\gamma t 
\alpha^{2}
\Delta^{\alpha-2}
q^{2}/(2\cos (\pi\alpha/2))=Q$, 
that is, 
\begin{eqnarray}
q & = & 
\sqrt{
\frac{2Q
\cos
\left(
\frac{\pi \alpha}{2}
\right)}
{
\gamma \alpha^{2}
\Delta^{\alpha-2}
}
}\, 
t^{1/2}, 
\label{eq:qToQ}
\end{eqnarray}
we obtain 
\begin{eqnarray}
P(t) & = & 
\psi 
(\alpha,\gamma,\Delta)\, 
t^{-3/2} \\
\psi (\alpha,\gamma,\Delta) & \equiv & 
\frac{\sqrt{2}}{\pi}
\left(
\frac{\gamma \alpha^{2}
\Delta^{\alpha-2}}
{
\cos
\left(
\frac{\pi \alpha}{2}
\right)
}
\right)^{-1/2}
\int_{\epsilon}^{\infty}
dS
\int_{0}^{\infty}
dQ\,
Q^{1/2}
{\rm e}^{Q}
\cos
\left[
\sqrt{
\frac{\gamma 
\alpha^{2}
\Delta^{\alpha-2}}
{
2\cos 
\left(
\frac{\pi \alpha}
{2}
\right)
}
}\,
Q^{1/2}S
\right].
\end{eqnarray}
It should be noted that 
this 
$t^{-3/2}$-law 
is valid for 
$|q| \leq \Delta$. 
From 
the equation 
(\ref{eq:qToQ}), 
this condition 
reads 
\begin{eqnarray}
t & \geq & 
\frac{2Q \cos 
\left(
\frac{\pi \alpha}{2}
\right)}
{
\gamma 
\alpha^{2}
\Delta^{\alpha}
} \equiv 
t_{\times}
\label{eq:cross_point}.
\end{eqnarray}
On the other hand, for 
$|q| > \Delta$, 
that is to say, 
for $t < t_{\times}$, 
the FPT distribution 
(\ref{eq:pt_TLF}) 
is 
evaluated as 
\begin{eqnarray}
P(t) & \simeq & 
-\frac{2\gamma}{\pi}
\int_{\epsilon}^{\infty}
dS
\int_{0}^{\infty}
dq 
\left[
\frac{q^{\alpha}
\left\{
\cos
\left(
\frac{\pi \alpha}{2}
\right) 
-
\left(
\frac{\Delta}{q}
\right)^{\alpha}
\right\}
}
{
\cos
\left(
\frac{\pi \alpha}{2}
\right)
}
\right] 
{\rm e}^{-\gamma |q|t}
\cos (qS) \nonumber \\
\mbox{} & \simeq & 
-\frac{2\gamma}{\pi}
\int_{\epsilon}^{\infty}
dS
\int_{0}^{\infty}
dq \,
|q|^{\alpha}\,
{\rm e}^{-\gamma |q|^{\alpha}t}
\cos (qS).
\end{eqnarray}
This result is 
identical to 
the FPT distribution 
for the 
conventional L$\acute{\rm e}$vy 
flight, which is 
defined by equations (\ref{eq:pt_Levy})
and (\ref{eq:pt_Levy2}), 
and was already obtained  
in the 
previous section. 

Let us 
summarise the result for 
the scaling laws of 
the FPT distribution 
for the truncated 
L$\acute{\rm e}$vy flight. 
\begin{eqnarray}
P(t) & \sim & 
\left\{
\begin{array}{ll}
t^{-(\alpha +1)/\alpha} & 
(t < t_{\times} : 
\mbox{non-Gaussian L$\acute{\rm e}$vy regime}) \\
t^{-3/2} & 
(t \geq t_{\times} :  
\mbox{Gaussian regime}) 
\end{array}
\right.
\end{eqnarray}
We should bear in mind that 
the 
crossover point 
$t_{\times}$ obtained 
by (\ref{eq:cross_point}) 
contains 
integral variable $Q$. 
Therefore, 
it is hard to say that 
$t_{\times}$ is 
well-defined. 
To delete 
the $Q$-dependence of 
$t_{\times}$, 
we consider the ratio 
of $t_{\times}(\Delta)$ and 
$t_{\times}(2\Delta)$. 
From equation (\ref{eq:cross_point}), 
we obtain  
$t_{\times}(\Delta)/t_{\times}(2\Delta)=2^{\alpha}$, 
namely, $t_{\times}(\Delta)=
2^{\alpha}t_{\times}(2\Delta)$. 
Let us 
check this 
scaling relation 
for the result we obtained 
in FIG. \ref{fig:fg1}. 
For $\alpha=1.2$, 
the relation 
reads 
$t_{\times}(\Delta)=
2^{1.2}t_{\times}(2\Delta)=
2.297 \,t_{\times}(2\Delta)$. 
This relation 
predicts the crossover point 
$t_{\times}(\Delta=0.5)/
t_{\times}(\Delta=1.0)=
t_{\times}(\Delta=1.0)/
t_{\times}(\Delta=2.0)=
2.297$, 
which is very close to 
the results obtained in 
FIG. \ref{fig:fg1}, 
namely, $44/19 \simeq 2.316$ and 
$19/8 \simeq 2.375$. 
The small difference is 
probably because of impreciseness of numerical integrations 
appearing in equation (\ref{eq:pt_TLF}).

The relation 
$t_{\times}(\Delta)=2^{\alpha}t_{\times}(2\Delta)$ 
for successive 
values of 
$\Delta$ and $2\Delta$ is easily extended 
for 
the relation between 
$\Delta$ and $\delta \Delta$ ($\delta \geq 1$) as follows. 
\begin{eqnarray}
t_{\times}(\Delta) & = & 
\delta^{\alpha}t_{\times}(\delta \Delta)
\label{eq:main_relation}
\end{eqnarray}
This scaling relation 
for the crossover point $t_{\times}$ 
in the scaling laws 
of the FPT distribution of 
the truncated L$\acute{\rm e}$vy flight 
$t_{\times}$ is 
one of the main 
results in this paper. 
From this result (\ref{eq:main_relation}), 
we find that 
the crossover point $t_{\times}$ 
increases rapidly as 
the effective cut-off length 
$l \equiv (\delta \Delta)^{-1}$ 
also increases as 
\begin{eqnarray}
t_{\times}(l) & = & 
\left(
\frac{l}{l_{0}}
\right)^{\alpha}\,
t_{\times}(l_{0})
\end{eqnarray}
where 
we set $l_{0} \equiv \Delta^{-1}$. 
Therefore, 
we conclude that 
the crossover between non-Gaussian 
L$\acute{\rm e}$vy and 
Gaussian 
regimes 
is 
observed 
not only in 
the distribution of 
the real-time $n$ flight, 
which 
was reported by 
Mantegna and Stanley \cite{Mantegna94}, 
but also in the 
FPT distribution 
of the truncated 
L$\acute{\rm e}$vy flight. 

In the study by Mantegna and Stanley \cite{Mantegna94}, 
they investigated 
the stochastic variable $z_{n}=\sum_{k=1}^{n}
x_{k}$, where 
$x_{k} \equiv X_{k}-X_{k-1}$ obeys the 
truncated L$\acute{\rm e}$vy flight. 
They evaluated the 
probability of return 
$P(z_{n}=0)$ 
and found that  
the $P(z_{n}=0)$ obeys 
the Gaussian $n^{-1/2}$-law 
in the large real time step $n$ regime. 
In this section, 
it was shown that this ultra-slow convergence 
from L$\acute{\rm e}$vy 
regime to the Gaussian regime 
is conserved even if we consider the 
first passage process of the 
truncated L$\acute{\rm e}$vy flight. 
The relation 
between their results and ours 
is clearly understood as follows. 

For a given 
time interval $t$ 
of the first passage 
process of 
the truncated L$\acute{\rm e}$vy flight, 
the time series 
of the variable $x_{k}=X_{k_{0}+k}-X_{k_{0}+k-1}$ 
behaves as 
$x_{1}, x_{2}, \cdots, x_{t}$, 
where $k_{0}$ is an origin for the measurement 
of the interval $t$. 
Then, from the observation 
by Mantegna and Stanley, 
the sum $z_{t}=\sum_{k=1}^{t}x_{k}=X_{k_{0}+t}-X_{k_{0}}$ 
obeys a Gaussian with zero-mean and variance $t$ if 
the time interval $t$ is large enough, 
that is, $t \geq t_{\times}$. 
Then, the probability of return 
is given by $P(z_{t}=0) \simeq 
t^{-1/2}$. 
In other words, 
for $t \geq t_{\times}$, it takes 
quite long time for a random-walker to escape from 
the region $[X_{k_{0}}-\epsilon, 
X_{k_{0}}+\epsilon]$, and 
the time $t$ for escaping 
guarantees that 
the central limit theorem works 
to make the variable 
$z_{t}$ a Gaussian. 
As a result, 
the FPT distribution $P(t)$ should 
follow the corresponding 
Gaussian $t^{-3/2}$-law 
from our argument 
for the case of the Wiener process (\ref{eq:result_Wiener2}). 
On the other hand, 
if the interval 
$t$ is smaller than 
the crossover point $t_{\times}$, 
the central limit theorem 
for $z_{t}$ 
does not work 
and $z_{t}$ is 
no longer a Gaussian. 
Then, as we checked, the FPT 
distribution $P(t)$ obeys $t^{-(\alpha +1)/\alpha}$-law 
of the L$\acute{\rm e}$vy flight. 
\section{Summary}
In this paper, 
we proposed 
a new 
approach to evaluate the FPT 
distribution. 
Our method is based on direct counting of the FPT. 
We show that our approach gives an explicit form of 
the FPT distribution for stable stochastic 
processes. 
Actually, for Wiener (Brownian motion), 
Lorentzian and L$\acute{\rm e}$vy 
stochastic processes, 
our method was demonstrated. 
Thanks to 
the mathematical simplicity of 
our method, 
it becomes easy to 
grasp the intuitive meaning 
of the FPT distribution and 
to tackle more complicated stochastic 
processes. 
As an example, 
we discussed 
the FPT distribution of 
the truncated L$\acute{\rm e}$vy flight (the KoBoL process). 
We found a clear crossover between 
non-Gaussian 
L$\acute{\rm e}$vy and Gaussian 
regimes 
in the scaling laws of the 
FPT distribution. 
We found the scaling relation 
on the crossover point $t_{\times}$ 
with respect to the effective 
length $l$ of 
the cut-off as 
$t_{\times} (l) =(l/l_{0})^{\alpha}t_{\times}(l_{0})$ with 
$l_{0}=\Delta^{-1}$. 

Very recently, Koren et.al. \cite{Koren} investigated 
not only the FPT distribution 
but also the first passage leapover (FPL) distribution 
under a single absorbing boundary condition. A 
relatively new concept, the FPL is defined 
as the flight length for a random walker to move beyond 
the single boundary (a target). Our system 
in this paper possesses two boundaries 
(in this sense, our process might 
be referred to as a first exit process); however, 
it might be possible to apply our analysis to 
the problem in order to discuss the FPL distribution. 
This will be addressed in future work. 

We hope that beyond the present 
analysis for the Sony Bank rate, our approach 
might be widely used 
in many scientific research 
fields, especially 
in the field of econophysics including financial data analysis. 
\begin{acknowledgments}
One of the authors (J.I.) was 
financially supported 
by {\it Grant-in-Aid for Young Scientists (B) 
of The Ministry of Education, Culture, 
Sports, Science and Technology (MEXT)} 
No. 15740229 and {\it Grant-in-Aid 
Scientific Research on Priority Areas 
``Deepening and Expansion of Statistical Mechanical Informatics (DEX-SMI)" 
of The Ministry of Education, Culture, 
Sports, Science and Technology (MEXT)} 
No. 18079001. 
N.S. would like to thank Shigeru Ishi, President 
of the Sony Bank, for useful discussions.
The authors thank the anonymous referee for many 
instructive comments on the manuscript. 
We also thank Enrico Scalas for fruitful 
discussion and useful comments. 
\end{acknowledgments}


\begin{thebibliography}{9999}
\bibitem{Redner}
S. Redner, {\it A Guide to First-Passage Processes}, 
Cambridge University Press (2001). 

\bibitem{Kappen}
N.G. van Kappen, 
{\it Stochastic Processes in 
Physics and Chemistry}, 
North Holland, Amsterdam (1992). 


\bibitem{Tuckwell}
H.C. Tuckwell, 
{\it Introduction to Theoretical 
Neurobiology}, 
Vol. 2, 
Cambridge University Press (1988). 

\bibitem{Tuckwell2}
H.C. Tuckwell, 
{\it 
Stochastic 
Processes in the Neurosciences}, 
Society for Industrial and Applied 
Mathematics, 
Philadelphia, Pennsylvania (1989). 

\bibitem{Simonsen}
I. Simonsen, 
M.H. Jensen and 
A. Johansen, 
Eur. Phys. J. B {\bf 27}, 
583 (2002). 

\bibitem{Raberto}
M. Raberto, 
E. Scalas and 
F. Mainardi, 
Physica A 
{\bf 314}, 749 (2002). 

\bibitem{Scalas}
E. Scalas, R. Gorenflo, H. Luckock, 
F. Mainardi, M. Mantelli and M. Raberto, 
Quantitative Finance 
{\bf 4}, 695 (2004). 



\bibitem{Kurihara}
S. Kurihara, 
T. Mizuno, 
H. Takayasu and 
M. Takayasu, 
{\it The Application 
of Econophysics}, 
H. Takayasu (Ed.), 
pp. 169-173, 
Springer (2003). 



\bibitem{Kaizoji}
T. Kaizoji and M. Kaizoji, 
Physica A {\bf 336}, 563 (2004).



\bibitem{Scalas06}
E. Scalas, Physica A {\bf 362}, 225 (2006).  



\bibitem{Sazuka}
N. Sazuka, 
Eur. Phys. J. B {\bf 50}, 129 (2006). 


\bibitem{Sazuka2}
N. Sazuka, Physica A {\bf 376}, 500 (2007).


\bibitem{SonyBank}
http://moneykit.net/


\bibitem{Rangarajan}
G. Rangarajan and 
M. Ding, Phys. Rev. E {\bf 62}, 120 (2000). 



\bibitem{Rangarajan2}
G. Rangarajan and M. Ding, 
Phys. Lett A. {\bf 273}, 322 (2000). 



\bibitem{Rangarajan3} 
G. Rangarajan and M. Ding, Fractals {\bf 8}, 139 (2000). 



\bibitem{Durrett}
R. Durrett, 
{\it Essentials of Stochastic Processes}, 
Springer-Verlag New York (1999). 


\bibitem{Chikara}
R. Chikara and L. Folks, 
{\it The Inverse Gaussian Density}, 
Dekker, New York (1989). 



\bibitem{Gerstner}
W. Gerstner and W. Kistler, 
{\it Spiking Neuron Models}, 
Cambridge University Press (2002). 


\bibitem{Mantegna94}
R.N. Mantegna and H.E. Stanley, 
Phys. Rev. Lett. 
{\bf 73}, 2946 (1994). 


\bibitem{Mantegna94_2}
R.N. Mantegna, 
Phys. Rev. E {\bf 49}, 
4677 (1994). 


\bibitem{Koponen}
I. Koponen, 
Phys. Rev. E {\bf 52}, 
1197 (1995). 



\bibitem{Mantegna2000}
R.N. Mantegna and H.E. Stanley, 
{\it An Introduction to 
Econophysics : 
Correlations and Complexity in 
Finance}, 
Cambridge University Press (2000). 


\bibitem{Risken}
H. Risken, 
{\it The Fokker-Plank Equation: Methods of 
Solution and Applications}, 
Springer-Verlag, Berlin; 
New York (1989). 


\bibitem{Gardiner}
C.W. Gardiner, 
{\it Handbook of Stochastic Methods}, 
Springer-Verlag, Berlin; New York (1983). 


\bibitem{Umeno}
K. Umeno, Phys. Rev. E {\bf 58}, 2644 (1998). 


\bibitem{KoBoL}
W. Schoutens, 
{\it L$\acute{\rm e}$vy Processes in 
Finance: Pricing Financial Derivatives}, 
Wiley, New York (2003). 


\bibitem{Boyarchenko}
S.I. Boyarchenko and 
S.Z. Levendorskii, 
{\it Generalizations of 
the Black-Scholes equation for 
truncated L$\acute{\rm e}$vy processes}, 
Working paper (1999). 


\bibitem{Voit}
J. Voit, 
{\it The Statistical Mechanics of Financial 
Markets}, Springer (2001). 



\bibitem{Koren}
T. Koren, A.V. Chechkin and J. Klafter, 
Physica A {\bf 379}, 10 (2007). 

\end{thebibliography}
\end{document}